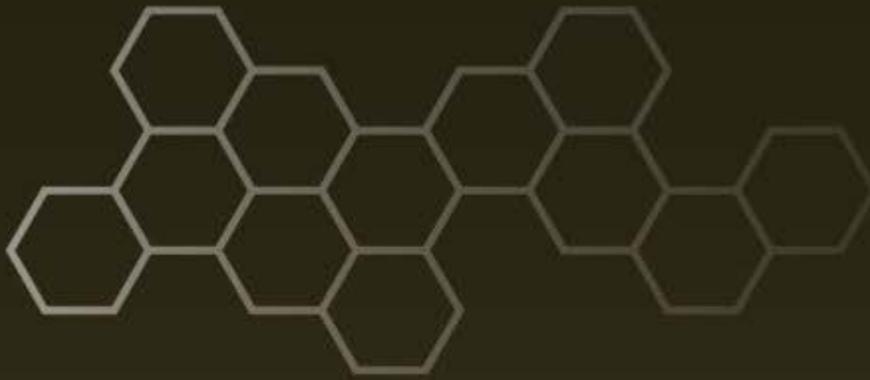
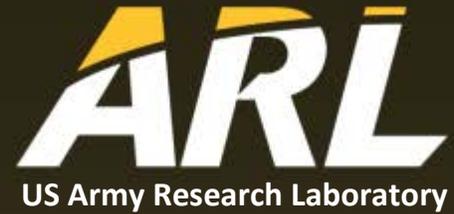

ARL-TR-7967 ● MAR 2017

**US Army Research Laboratory**

# Recommendations for Model-Driven Paradigms for Integrated Approaches to Cyber Defense

by Mona Lange, Alexander Kott, Noam Ben-Asher, Wim Mees, Nazife Baykal, Cristian-Mihai Vidu, Matteo Merialdo, Marek Malowidzki, and Bhopinder K Madahar

**NOTICES**

**Disclaimers**

The findings in this report are not to be construed as an official Department of the Army position unless so designated by other authorized documents.

Citation of manufacturer's or trade names does not constitute an official endorsement or approval of the use thereof.

Destroy this report when it is no longer needed. Do not return it to the originator.

ARL-TR-7967 ● MAR 2017

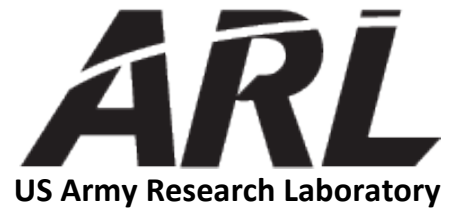

US Army Research Laboratory

# Recommendations for Model-Driven Paradigms for Integrated Approaches to Cyber Defense

by Mona Lange
*University of Lübeck, Germany*

Alexander Kott and Noam Ben-Asher
*Computational and Information Sciences Directorate, ARL*

Wim Mees
*Royal Military Academy, Belgium*

Nazife Baykal
*Middle Eastern Technical University, Turkey*

*Cristian-Mihai Vidu*
*National University for Political Sciences and Public Administration, Romania*

*Matteo Merialdo*
*Roma Tre University, Italy*

*Marek Malowidzki*
*Military Communication Institute, Poland*

*Bhopinder K Madahar*
*Defence Science and Technology Laboratory, United Kingdom*

| REPORT DOCUMENTATION PAGE | | | | | Form Approved<br>OMB No. 0704-0188 |
|---|---|---|---|---|---|
| Public reporting burden for this collection of information is estimated to average 1 hour per response, including the time for reviewing instructions, searching existing data sources, gathering and maintaining the data needed, and completing and reviewing the collection information. Send comments regarding this burden estimate or any other aspect of this collection of information, including suggestions for reducing the burden, to Department of Defense, Washington Headquarters Services, Directorate for Information Operations and Reports (0704-0188), 1215 Jefferson Davis Highway, Suite 1204, Arlington, VA 22202-4302. Respondents should be aware that notwithstanding any other provision of law, no person shall be subject to any penalty for failing to comply with a collection of information if it does not display a currently valid OMB control number.<br>**PLEASE DO NOT RETURN YOUR FORM TO THE ABOVE ADDRESS.** | | | | | |
| **1. REPORT DATE** (DD-MM-YYYY)<br>March 2017 | | **2. REPORT TYPE**<br>Technical Report | | **3. DATES COVERED** (From - To)<br>1 January 2016–January 2017 | |
| **4. TITLE AND SUBTITLE**<br>Recommendations for Model-Driven Paradigms for Integrated Approaches to Cyber Defense | | | | **5a. CONTRACT NUMBER** | |
| | | | | **5b. GRANT NUMBER** | |
| | | | | **5c. PROGRAM ELEMENT NUMBER** | |
| **6. AUTHOR(S)**<br>Mona Lange, Alexander Kott, Noam Ben-Asher, Wim Mees, Nazife Baykal, Cristian-Mihai Vidu, Matteo Merialdo, Marek Malowidzki, Bhopinder K Madahar | | | | **5d. PROJECT NUMBER** | |
| | | | | **5e. TASK NUMBER** | |
| | | | | **5f. WORK UNIT NUMBER** | |
| **7. PERFORMING ORGANIZATION NAME(S) AND ADDRESS(ES)**<br>US Army Research Laboratory<br>ATTN: RDRL-CIN<br>2800 Powder Mill Road<br>Adelphi, MD 20783-1138 | | | | **8. PERFORMING ORGANIZATION REPORT NUMBER**<br>ARL-TR-7967 | |
| **9. SPONSORING/MONITORING AGENCY NAME(S) AND ADDRESS(ES)** | | | | **10. SPONSOR/MONITOR'S ACRONYM(S)** | |
| | | | | **11. SPONSOR/MONITOR'S REPORT NUMBER(S)** | |
| **12. DISTRIBUTION/AVAILABILITY STATEMENT**<br>Approved for public release; distribution unlimited. | | | | | |
| **13. SUPPLEMENTARY NOTES** | | | | | |
| **14. ABSTRACT**<br>This report describes the findings of the North Atlantic Treaty Organization Exploratory Team investigating cyber defense. Many defensive activities in cyber warfare and information assurance rely on ad hoc techniques. The cyber community recognizes that comprehensive, systematic, principle-based modeling and simulation are more likely to produce long-term, reusable approaches. A model-driven paradigm is predicated on mechanisms of modeling the organization whose mission is under cyber attack, the mission itself, and the systems that support it. The level of detail of this class of problems ranges from the level of host and network events to systems assets and up to business functions. Solving this class of problems is of significant difficulty. Such modeling could be used to explore multiple alternative mitigation strategies and select optimal mitigating actions. The paradigm applied to cyber operations is likely to benefit traditional disciplines of cyber defense. The team identified challenges for model-driven paradigms for cyber defense and reviews 2 in detail: 1) modeling adversarial aspects, including wargaming, of the cyber warfare, and 2) modeling of human cognitive processes in relation to cyber activities. Based on its discussions, the team makes recommendations on modeling and simulation for a broad range of cyber defense disciplines. | | | | | |
| **15. SUBJECT TERMS**<br>cyber modeling, cyber defense, wargaming, modeling and simulation, risk, likelihood and impact models, adversary models, human cognition, resilience engineering, graph theory | | | | | |
| **16. SECURITY CLASSIFICATION OF:** | | | **17. LIMITATION OF ABSTRACT** | **18. NUMBER OF PAGES** | **19a. NAME OF RESPONSIBLE PERSON**<br>Alexander Kott |
| **a. REPORT**<br>Unclassified | **b. ABSTRACT**<br>Unclassified | **c. THIS PAGE**<br>Unclassified | UU | 58 | **19b. TELEPHONE NUMBER** (Include area code)<br>301-394-1507 |

Standard Form 298 (Rev. 8/98)
Prescribed by ANSI Std. Z39.18

# Contents

## List of Figures

INTENTIONALLY LEFT BLANK.

## Executive Summary

The North Atlantic Treaty Organization (NATO) Exploratory Team meeting, "Model-Driven Paradigms for Integrated Approaches to Cyber Defense", was organized by the NATO Science and Technology Organizations' (STOs') Information Systems and Technology (IST) panel and conducted its meetings and electronic exchanges during 2016. This report describes the proceedings and outcomes of the team's efforts.

Many of the defensive activities in the fields of cyber warfare and information assurance rely on essentially ad hoc techniques. The cyber community recognizes that comprehensive, systematic, principle-based modeling and simulation are more likely to produce long-term, lasting, reusable approaches to defensive cyber operations.

A model-driven paradigm is predicated on creation and validation of mechanisms of modeling the organization whose mission is subject to assessment, the mission (or missions) itself, and the cyber-vulnerable systems that support the mission. This by any definition is a complex socio-technical system (of systems), and the level of detail of this class of problems ranges from the level of host and network events to the systems' functions up to the function of the enterprise/business. Solving this class of problems is of medium to high difficulty and can draw in part on advances in Systems Engineering (SE). Such model-based approaches and analysis could be used to explore multiple alternative mitigation and work-around strategies and to select the optimal course of mitigating actions. Furthermore, the model-driven paradigm applied to cyber operations is likely to benefit traditional disciplines of cyber defense such as security, vulnerability analysis, intrusion prevention, intrusion detection, analysis, forensics, attribution, and recovery.

The team identified a number of challenges for model-driven paradigms for cyber defense and elected to review 2 of them in detail: the problems of modeling the adversarial aspects, including wargaming, of cyber warfare and modeling human cognitive processes in relation to cyber activities.

Recommendations include the following:

- Stress the need for modeling and simulation for full range of cyber specialties, not only for training and rehearsal.
- Encourage participation of commercial companies, in NATO STO activities and meetings, with an option to demonstrate their relevant products.

- Simulation models that we call here "Business Impact Simulation" are particularly important for NATO, but simulation models we call here "Attack Details Simulation" are crucial for NATO.

- Produce a set of clear and concrete requirements for modeling and simulation (M&S) tools specifically targeted at cyber defense and leveraging advances in SE.

- Simulation of attack-defense scenarios at a level of observable, component-network-and-system-level events.

- Minimize government investments in the line of approaches based on attack graphs and related methods to invest in other directions of cyber (M&S).

- Encourage academic research targeted at effective and validated M&S of human cognitive processes and behaviors as they execute cyber defense and attack.

The team initiated publication of a special issue of the *Journal of Defense Modeling and Simulation* dedicated specifically to model-driven paradigms for cyber defense. It also formulated a Technical Activity Proposal and obtained NATO IST approval for a workshop titled "Modelling and Simulation S&T: Critical Enabler for Cyber Defence", details of which appear in Appendix A.

# 1. Introduction

This report describes the proceedings and outcomes of the North Atlantic Treaty Organization (NATO) Exploratory Team meeting, "Model-Driven Paradigms for Integrated Approaches to Cyber Defense" (IST-ET-094), organized by the NATO Science and Technology Organizations' (STOs') Information Systems and Technology (IST) panel. Two meetings for IST-ET-094 were held: an inaugural workshop at the University of Lübeck in Lübeck, Germany, 14–18 March 2016, and the final meeting at the Royal Military Academy in Brussels, Belgium, 12–14 September 2016.

The STO's mission is to help position the NATO nations' and NATO's science and technology (S&T) investments as a strategic enabler of the knowledge and technology advantage for the defense and security posture of NATO nations and partner nations. This is accomplished by conducting and promoting S&T activities that augment and leverage the capabilities and programs of the alliance, of the NATO nations, and the partner nations, in support of NATO's objectives. It is further accomplished by contributing to NATO's ability to enable and influence security and defense-related capability development and threat mitigation in NATO nations and partner nations, in accordance with NATO policies, and by supporting decision making in the NATO nations and NATO.

IST, the immediate sponsor of this workshop, is one of the 5 NATO S&T panels whose role it is, with the NATO modeling and simulation (M&S) Group, to implement, on behalf of the S&T Board, the STO mission with respect to information systems technology. The focus of this panel is the advancement and exchange of techniques and technologies to provide timely, affordable, dependable, secure, and relevant information to warfighters, planners, and strategists, as well as enabling information systems technologies for modeling, simulation, and training. The IST covers the fields of information warfare and assurance, architecture and intelligent information systems, and communications and networks.

The motivation for the workshop had to do with the fact that while many of these defensive activities currently rely on essentially ad hoc techniques, there is a growing realization within the cyber community that comprehensive, systematic, principle-based M&S are more likely to produce long-term, lasting, reusable approaches to defensive cyber operations. In particular, a recent NATO workshop (IST-128-RWS-019, "Cyber Attack Detection, Forensics and Attribution for Assessment of Mission Impact") voiced a strong consensus that substantive solutions for at least some classes of cyber challenges, for example mission impact assessment, require adopting and developing a new model-driven paradigm. Such a paradigm is predicated on creation and validation of mechanisms of modeling the organization whose mission is subject to assessment, the mission (or missions) itself, and the cyber-vulnerable systems that support the mission.

The models are then used to simulate or otherwise portray the cyber attacks and associate defensive phenomena and system operations, including assessment of mission impact. This by any definition is a complex socio-technical system (of systems) and the level of detail of this class of problems ranges from the level of host and network events to the systems functions up to the function of the enterprise/business. Solving this class of problems is of medium to high difficulty and can draw in part on advances in Systems Engineering (SE). Such model-based approach and analysis could be used to explore multiple alternative mitigation and work-around strategies and then select the optimal course of mitigating actions. Furthermore, the model-driven paradigm applied to cyber operations is likely to benefit traditional disciplines of cyber defense, such as security, vulnerability analysis, intrusion prevention, intrusion detection, analysis, forensics, attribution, and recovery. For example, intrusion detection, especially for zero-day or polymorphic attacks, would greatly benefit from the ability to model the observable effects of a hypothetical attack.

The team identified 2 major subspaces of model-driven paradigms for attack analysis cyber modeling and simulation problems in attack analysis, as shown in Fig. 1.

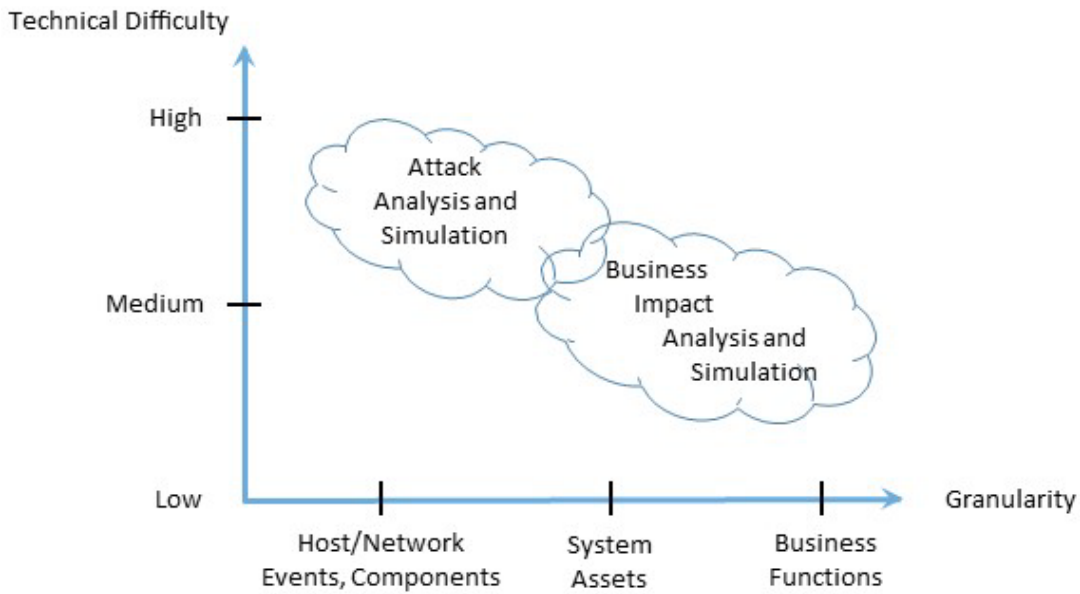

**Fig. 1**     **Two major subspaces of cyber modeling and simulation problems**

The team identified a number of challenges for model-driven paradigms for cyber defense and elected to review 2 of them in detail. A section of this report reviews the problem of modeling the adversarial aspects, including wargaming, of cyber warfare. Another section reviews the modeling of human cognitive processes in relation to cyber activities.

Other results and outcomes of the Exploratory Team included the following:

- The team formulated a number of conclusions and recommendations, summarized in Section 5 of this report.

- The team initiated publication of a special issue of the *Journal of Defense Modeling and Simulation* dedicated specifically to model-driven paradigms for cyber defense (Appendix B).

- The team formulated a Technical Activity Proposal and obtained NATO IST approval for a workshop titled "Modelling and Simulation S&T: Critical Enabler for Cyber Defence" (Appendix C).

- The team explored a number of commercial and open-source tools that appeared to be relevant to modeling and simulation of cyber security activities (Appendix D).

- The team started to explore model-based approaches in SE in terms of utility for cyber defense, the rationale being that such approaches are gaining commercial use for the design and development of information technology systems and components and therefore are relevant to cyber M&S. As the broader developments and thinking in this area remained immature, it was identified as a topic of specific interest in the workshop (see Appendix C).

## 2. Contributors

Participating nations included Belgium, Bulgaria, Germany, the United Kingdom (UK), Poland, Romania, Turkey, and the United States. The participants included the following:

- Alexander Kott (USA)
- Wim Mees (Belgium)
- Felix Kuhr (Germany)
- Nazife Baykal (Turkey)
- Cristian-Mihai Vidu (Romania)

Approved for public release; distribution unlimited.

- Matteo Merialdo (Italy)
- Mona Lange (Germany)
- Marek Malowidzki (Poland)
- Bhopinder K Madahar (UK)

Remote (teleconference) attendees included the following:

- Margaret Varga (UK)
- Nikolai Stoianov (Bulgaria)

## 3. Modeling Adversarial Aspects of Cyber Attacks

Primary Author: Alexander Kott (USA)

The ability to reason about or model actions of the adversary in relation to defender's actions appears to be a significant aspect of modeling in cyber defense. Terms like "wargaming" or "adversarial reasoning" apply to such forms of M&S. In this section, we review the literature and provide related observations regarding these aspects of modeling in cyber defense.

**An Illustrative Description of a Class of Modeling with Wargaming**

Exploring wargaming aspects of cyber modeling, ET-094 elected to focus on a specific class of problems. In the following, we describe the class that has to do with in-depth analysis of cyberattack progression inspired by adversarial reasoning and automated wargaming approaches in conventional, kinetic warfare (e.g., Kott et al. 2002; Rasch et al. 2003).

Given data about friendly infrastructure, applications, and defensive mechanisms, and some assumptions about enemy intent and capability, the objective was to find the following:

- A set of most-probable detailed steps of enemy courses of action (COAs), as modulated by friendly defensive actions
- Detailed steps should be at observable level of details such as could be found in host system logs and network traffic packet capture files.
- For each COA estimate, find the probability of enemy actions failing or detected by friendly systems/people.
- Include considerations of cognitive processes and limits of defenders (and perhaps of attackers)

- Target this type of model for several types of integrated utilization: developing prevention and detection agents, forensics, risk assessment, friendly COA development, and cyber battle command.

## 3.1 Taxonomy and Literature Review of Common Types of Models in Cyber Defense

The following comprise an overview of the taxonomy and definitions used for the review:

- Emulation (often with simulation) of networks: actual hardware, software, and humans (e.g., cyber ranges)

- Training-focused simulations: presenting to human trainees the effects of a cyber attack without modeling underlying processes

- M&S of human cognitive processing of cyber events and situations: perception, recognition, situational awareness (SA), and decision making

- M&S of attack progress and malware propagation

    o Attack-graph-based approaches

    o Epidemiology analogy (e.g., Susceptible, Infected, Recovered [SIR])

- Abstract wargaming: game-theoretic model of cyber conflict without modeling the underlying processes of cyber attack and defense

- Business processes models: defense, offense, and business processes, along with business information technology architecture, simulated for observing resulting effects

- Statistical models of cyber events: cyber processes represented as, for example, equations of Poisson processes, and coefficients learned from training dataset

- Two classes of models used to support cyber modeling but do not model cyber aspects:

    - Physical systems models to support modeling of cyber-physical effects

    - Network simulation models

The following section summarizes each element of the taxonomy with brief review of related literature.

## 3.2 Emulation

This often includes elements of both emulation and simulation of networks, simulation of an attack on the network and its progress, and might include human defenders and attackers. Many cyber ranges (CRs) are based on this paradigm, which is often used to evaluate a system or technology concept, as evidenced in the following literature:

- CyberVAN [accessed 23 Jan 2016]. http://www.appcomsci.com/research/tools/cybervan.

- Serban C et al. Testing android devices for tactical networks: a hybrid emulation testbed approach. Proceedings of the Military Communications Conference. New York (NY): Institute of Electrical and Electronics Engineers; 2015.

   Davis J, Magrath S. A survey of cyber ranges and testbeds. Edinburgh (Australia): Cyber and Electronic Warfare Division, Defence Science and Technology Organization; 2013. Report No.: DSTO-GD-0771.

   The authors explored the approaches used to build existing CRs, the merits of each approach and their functionality. The review first categorizes CRs by their type and second by their supporting sector: academic, military, or commercial. The types of CR are identified as simulation, overlay, or emulation. CRs are considered simulations if they use software models of real world objects to explore behavior. They are labelled as overlays if they operate on live production hardware with experiments sharing their production resources rather than using a dedicated CR laboratory. CRs are labelled emulations if they run real software applications on dedicated hardware. Emulation refers to the software layer that allows fixed CR hardware to be reconfigured to different topologies for each experiment. The review found CRs are predominantly used for training.

In ET-094 foci, humans are also simulated for purposes of reducing bias, improving controllability, repeatability of runs, and the ability to perform large number of runs to discover novel patters and phenomena. Thus, we focus on full constructive simulations.

Such models are well suited for wargames between human attackers and defenders. Unfortunately, repeatability of such wargames is poor, and it is difficult to make generalized conclusions from such wargames.

### 3.3 Training-Focused Simulations

These are usually limited to presenting to the trainee with the effects of a cyber attack without modeling underlying processes. Examples in literature include the following:

- Marshall H et al. Cyber operations battlefield web services (COBWebS): concept for a tactical cyber warfare effect training prototype. Proceedings of the 2015 Fall Simulation Interoperability Workshop; Simulation Interoperability Standards Organization (SISO): 2015.

- Littlejohn AM, Makhlouf E. Test and evaluation of the malicious activity simulation tool (MAST) in a local area network (LAN) running the common PC operating system environment (COMPOSE). Monterey (CA): Naval Postgraduate School; 2013.

  The paper notes that training network administrators is a rather expensive and time-consuming process. MAST aims to provide realistic, tailored simulation of malicious activity for the purpose of training network administrators to recognize and respond to threats on the network they manage.

- Chapman IM et al. Taxonomy of cyber attacks and simulation of their effects. Proceedings of the 2011 Military Modeling & Simulation Symposium. San Diego (CA): Society for Computer Simulation International; 2011.

  The authors propose methods to simulate the effects of several cyber attack types for use in simulation in support of training and experimentation.

Such models are not generally suitable for wargaming.

### 3.4 M&S of Human Cognitive Processing

These focus on modeling how humans process cyber events and situations: perception, recognition, SA, and decision making, examples of which are in the following literature:

- Cho J-H, Cam H, Oltramari A. Effect of personality traits on trust and risk to phishing vulnerability: modeling and analysis. Cognitive Methods in Situation Awareness and Decision Support (CogSIMA), IEEE International Multi-Disciplinary Conference; 2016 Nov 2–4; Beirut, Lebanon.

- Ben-Asher N, Oltramari A, Erbacher RF, Gonzalez C. Ontology-based adaptive systems of cyber defense. Proceedings of the 10th International Conference on Semantic Technology for Intelligence, Defense, and Security (STIDS); 2015.

  Jones RM, et al. Modeling and integrating cognitive agents within the emerging cyber domain. Proceedings of the Interservice/Industry Training, Simulation, and Education Conference (I/ITSEC); 2015.

  The application of cognitive agents to the roles of cyber offense and defense would provide a more complete cyber ecology for training purposes and thus a more realistic training experience for the student. There are 2 key challenges to creating such cyber agents: 1) modeling the complex and continually evolving processes of cyber operations within a cognitive architecture and 2) defining the tools and data standards to enable cognitive agents to interoperate with networks in a portable way. This paper discusses novel models of cyber offensive and defensive behavior based on observation and elaboration of human expertise. It also offers an approach to the creation of software adapters that translate from task-level actions to network-level events to support agent-network interoperability.

- Robinson D, Cybenko G. A cyber-based behavioral model. The Journal of Defense Modeling and Simulation: Applications, Methodology, Technology. 2012;9(3):195–203.

  This research examines aspects of the cognitive level by defining a cyber-based behavioral model contingent on the activities a user performs while on the Internet. The authors believe this is foundational to completely define a cyber SA model, thus providing commanders and decision makers a more comprehensive and real-time view of the environment in which they are operating.

Such models are not directly suitable for wargaming but can be valuable components of a complete wargaming system.

### 3.5 M&S Based on Attack Graph Concept

Examples in literature include the following:

- Skybox Security. Using risk modeling and attack simulation for proactive cyber security [accessed 23 Jan 2016]. https://www.skyboxsecurity.com/.

  Explains what they mean by cyber M&S and its purposes. Their approach is clearly not able to reproduce observables, and therefore they do not mention detection and forensics as its purposes, but they do stress predictive aspect.

- Kotenko I, Chechulin A. A cyber attack modeling and impact assessment framework. Proceedings of the 5th International Conference on Cyber Conflict. New York (NY): Institute of Electrical and Electronics Engineers; 2013.

  The paper suggests a framework for cyber attack modeling and impact assessment. It is supposed that the common approach to attack modeling and impact assessment is based on representing malefactors' behavior, generating attack graphs, calculating security metrics, and providing risk analysis procedures. The architecture of the Cyber Attack Modeling and Impact Assessment Component (CAMIAC) is proposed. The authors present the prototype of the component, the results of experiments carried out, and comparative analysis of the techniques used.

- Sommestad T et al. The cyber security modeling language: a tool for assessing the vulnerability of enterprise system architectures. IEEE Systems Journal. 2013;7(3):363–373.

  The Cyber Security Modeling Language (CySeMoL) is a modeling language for enterprise-level system architectures coupled to a probabilistic inference engine. If the computer systems of an enterprise are modeled with CySeMoL, this inference engine can assess the probability that attacks on the systems will succeed. The theory used for the attack-probability calculations in CySeMoL is a compilation of research results on a number of security domains and covers a range of attacks and countermeasures.

Such models are not normally used for wargaming purposes. In principle, a wargaming element could be added if the defender could execute actions directed at closing some attack paths.

### 3.6 M&S Based on Epidemiology Analogy

These use the popular SIR-type models. Examples in the literature include the following:

- Marvel LM et al. A framework to evaluate cyber agility. Proceedings of the 2015 Military Communications Conference. New York (NY): Institute of Electrical and Electronic Engineers (IEEE); 2015.

- Thompson B, Morris-King J, Cam H. Controlling risk of data exfiltration in cyber networks due to stealthy propagating malware; 2017 in press.

Using a SIR-type approach, the authors present a compartmental stochastic model to represent changes in the state of the network and analytically derive an upper bound on the rate at which an attacker can exfiltrate data without arousing suspicion.

Such models are not normally used for wargaming. However, they could be if defender could execute actions that limit the spread of malware.

### 3.7 Abstract Wargaming

Here, a game-theoretic process is modeled with moves and effects inspired by cyber conflict but without modeling the underlying processes of cyber attack and defense. Examples in literature include the following:

- Cho J-H, Gao J. Cyber war game in temporal networks. PloS One. 2016;11(2);e0148674.

- Manshaei MH et al. Game theory meets network security and privacy. ACM Computing Surveys (CSUR). 2013;45(3):25.

    This survey provides a structured and comprehensive overview of research on security and privacy in computer and communication networks that uses game-theoretic approaches. The authors organize the presented works in 6 main categories: 1) security of the physical and media access control layers, 2) security of self-organizing networks, 3) intrusion detection systems, 4) anonymity and privacy, 5) economics of network security, and 6) cryptography. In each category, they identify security problems, players, and game models.

- Backhaus S et al. Cyber-physical security: a game theory model of humans interacting over control systems. IEEE Transactions on Smart Grid. 2013;4(4):2320–2327.

    Accurate predictions require good models of not just the physical and control systems, but also of human decision making. In this article, the authors present an approach to develop such tools, specifically models of the decisions of the cyber-physical intruder who is attacking the systems and the system operator who is defending it, and demonstrate its usefulness for design.

Such models are explicitly designed for game-solving purposes. It is not exactly wargaming but very close.

## 3.8 Business Process-Based Models

Defense, offense, and business processes are modeled as process diagrams and simulated for the sake of observing resulting effects. An example in the literature is the following:

- Noel S et al. Analyzing mission impacts of cyber actions (AMICA). NATO; 2015. Report No.: IST-128.

Such models are essentially used for automated wargaming.

## 3.9 Statistical Models

Here, cyber processes are represented as, for example, equations of Poisson process and coefficients are learned from training dataset, as in the following example:

- Gil S, Kott A, Barabási A-L. A genetic epidemiology approach to cyber-security. Scientific Reports 4. 2014 July 16. doi:10:1038/srep05659.

By themselves, such models are not suitable for wargaming. However, they could be a useful component of a wargaming system.

## 3.10 Maturity Models

There is growing research interest in developing models that can better represent the ability and maturity of organizations to manage risks and threats in cyber space. These are conceptual models, building on software engineering and information technology (IT) risk management approaches and others, to outline and assess characteristics associated with various levels of maturity. Consequently they serve as weakness indicators and mitigation/improvement maps to help improve cyber defenses and resilience. Examples in the literature include the following:

> Poeppelbuss J et al. Maturity models in information systems research: literature search and analysis. Communications of the Association for Information Systems. 2011;29(1).
> 
> Reviewed research related to models such as Capability Maturity Model (e.g., CMM and CMMi), which are well known. Findings are that research was generally critical to the applicability and reliability of existing models, which need greater rigor in their measurement and conceptualization.

- Becker J et al. Developing maturity models for IT management – a procedure model and its application. Business & Information Systems Engineering. 2009;3.

- Though the focus is on the development of the IT Performance Measurement Maturity Model, the research identifies the gaps that need to be addressed. It uses the guidelines for Design Science proposed by Hevner (2004) and postulates 8 requirements for the design process of models. This opens up potential for improvements (e.g., detailed specifications and epistemological substantiation of requirements) and inform higher-level cyber maturity models.

- White GB. The community cyber security maturity model. Proceedings of the 40th Hawaii International Conference on System Sciences; 2007.

  This paper considers the development of maturity models for improving cyber defense within communities; that is, assessment of preparedness against cyber attacks. A Community Cyber Security Maturity Model is proposed to identify weaknesses and develop programs to enhance security posture.

- Healey J, van Bochoven L. NATO's cyber capabilities: yesterday, today, and tomorrow. Atlantic Council, Smarter Alliance Initiative Issue Brief; 2012.

  This is a think piece on NATO cyber capabilities and recommendations as to the key enablers for the future. One of these is adoption of relevant information security and IT standards and adoption of models related to risk and maturity such as the Resilience Management Model.

These works suggest linking developments across security, IT risk management, and business resilience areas for assessing and enhancing cyber capability of organizations. A software-based framework, the Cyber Defense Capability Assessment Tool (see https://apmg-cyber.com/products/cdcat), has been developed to model cyber capabilities of an enterprise. Though such frameworks and maturity models are not currently part of cyber M&S, they do provide an important contextual view within which the more detailed M&S can be undertaken; for example, by identifying where the cyber weaknesses are within an enterprise (e.g., people, processes, and technology) and applying M&S to specific parts to exercise higher-level decisions and actions.

Maturity models are not suitable for wargaming but can be used to provide context to exercises such as exploring maturity of processes.

## 4. Human Cognitive Modeling in Cyber Security

Primary Author: Noam Ben-Asher (USA)

Many technological advances in cyber security have facilitated advanced monitoring and threat detection. However, the ever-growing sophistication of attackers deters the ongoing attempt to automate critical tasks that cyber defenders perform. Human analytical capabilities, the ability to discern suspicious activity, and authority to make decisions on threats place the human defender as a key player in cyber defense. As such, there is a substantial interest in understanding human cognitive activities and behaviors that drive detection of cyber threats and response selection. Modeling and simulating these processes can allow us to discern the underlying mechanisms, revealing their strengths and deficiencies. Modeling and simulating also provide the necessary conditions for proactive defense.

However, progression in modeling and simulating human cognition in cyber defense confronts significant hurdles. The first relates to the limited availability of reliable data that capture defenders' decision making processes as well as the cyber environment in which they operate. Besides it being extremely hard to collect highly complex data, in many cases such data are confidential and cannot be shared with the research community. A second hurdle relates to the lack of unified theory of cyber detection in general and specifically defining the role defenders' cognition plays, what mental models defenders use, and how human cognition interfaces with detection tools. As a result, research on cyber defenders' cognitive activities and behaviors is highly fragmented. Furthermore, due to the complexity of the domain and the limited ability to collect data from professional security experts, the tendency to use simplified tasks and experimentation often involves participants with limited or no understanding in cyber security.

Gonzalez et al. (2014) identified the following 4 critical aspects of cyber defender cognitive modeling:

- High levels models of defender-in-the-loop
- Perception and representation of the highly dynamic cyber environment, including vulnerabilities, assets, and threats among others
- Cyber SA, how it relates to SA in the physical world, and what are the cognitive demands to achieve cyber SA
- Decision-making processes including learning, adaptation, and decision biases

- Scaling up from the individual defender to interactions within a team of defenders or between a defender and an attacker

In the following we review literature related to each of these domains.

## 4.1 Models of Human in the Cyber Detection Loop

High-level modeling of defender-in-the-loop often attempts to map and define the human analyst role in detection processes. These high-level models often focus on information flows, task analysis, and interactions with decision support tools. Such preliminary models can provide systematic mapping of the roles, tasks, and responsibilities of human defenders, plus what tools are used in different scenarios, and lay the foundations for automating some of these tasks. Examples in literature include the following:

- D'Amico A, Whitley K. The real work of computer network defense analysts. Proceedings of the 4th International Workshop on Visualization for Cyber Security. Berlin (Germany): Springer; 2008. p. 19–37.

- D'Amico A et al. Achieving cyber defense SA: a cognitive task analysis of information assurance analysts. Proceedings of the Human Factors and Ergonomics Society Annual Meeting. 2005;49(3).

    One of the most human labor-intensive aspects of cyber defense is triage analysis, where the human defenders review and sort a large volume of network activities data that already happened and corresponding intrusion detection alerts. These studies are based on cognitive task analysis conducted to baseline the state of the practice in the US Department of Defense Computer Network Defense community. Based upon these observations, the authors propose an analyst-in-the-loop workflow model that encapsulates analytic goals, tasks, types of decisions made, data sources used to make those decisions, cognitive demands, and tools used.

- Oltramari A et al. General requirements of a hybrid-modeling framework for cyber security. Proceedings of the 2014 IEEE Military Communications Conference; 2014.

- Jones RM et al. Modeling and integrating cognitive agents within the emerging cyber domain. Proceedings of the Interservice/Industry Training, Simulation, and Education Conference (I/ITSEC); 2015.

    The authors propose to harness cognitive architectures such as Soar and ACT-R to build executable cognitive agents that can interact with real cyber networks in real time. This all-embracing development of a cognitive model

of a cyber defender is driven by cognitive task analysis and knowledge elicitation from subject matter experts. This knowledge is assimilated into the procedural, semantic, and episodic long-term memories of the cognitive agent. Based on goals and inputs from perception, an agent can query long-term memories and choose the appropriate actions. Observing the outcomes of an action and changes in the environment closes the learning loop of the agent. While there are constant advancements in the understanding of cognitive processes and the maturity of cognitive architecture also increase, the prospect of developing a generic cyber defender using this approach in the near future is fairly limited. This is mainly due to difficulty in successfully capturing and assimilating all of the required knowledge regarding the complex dynamics of the cyber environment. It is more likely to benefit from small-scale specialized cognitive agents.

- Zhong C et al. Automate cybersecurity data triage by leveraging human analysts' cognitive process. Presented at the 2nd Big Data Security on Cloud; 2016 Apr 8–10; Columbia University, NY.

  The authors demonstrate how by capturing and analyzing traces of human defender's behavior, it is possible to construct a model of the defender's decision making process. A cognitive agent based on this model is then used to automate triage analysis yielding faster performance than human defenders but with similar accuracy.

- Greitzer FL, Hohimer RE. Modeling human behavior to anticipate insider attacks. Journal of Strategic Security. 2011;4(2):25.

  This study combines traditional cyber security audit data with psychosocial data to develop a predictive model that supports the detection of insider threats. It reports on progress in defining a usable set of predictive indicators and developing a framework for integrating the analysis of organizational and cyber security data to yield predictions about possible insider exploits and developing the knowledge base and reasoning capability of such a system. The authors also outline the types of errors that one expects in a predictive system versus a detection system and discuss how those errors can affect the usefulness of their model.

## 4.2 Perception and Representation of the Cyber Environment

Cyber environments in which defenders operate are highly dynamic and differ considerably between organizations. As such, most cyber environments challenge human analytical capabilities by including high volume of data that come from a variety of sensors. These real-time, high-velocity data streams with inherent veracity serve as the basis for the defender's SA and decision-making processes. Therefore, multiple studies investigate and model how defenders perceive and internally represent cyber environments, including the following:

- Ben-Asher N, Oltramari A, Erbacher RF, Gonzalez C. Ontology-based adaptive systems of cyber defense. Proceedings of the 10th International Conference on Semantic Technology for Intelligence, Defense, and Security (STIDS); 2015.

  Among other things, successful modeling of human cognition in cyber defense must carefully attend human knowledge representation and the interplay between knowledge representation and decision making. The authors use computational ontology to provide a knowledge base for an adaptive cognitive agent and evaluate the ability of this hybrid architecture to detect malicious network activity. The detailed representation of the cyber environment allowed a relatively simple cognitive model of a defender to learn from experience and accurately detect malicious port-scanning. Beyond emphasizing the key role of in-depth knowledge representation in modeling human cognition, the paper demonstrates how knowledge is used in decision making, learning, and detection.

- Robinson D, Cybenko G. A cyber-based behavioral model. The Journal of Defense Modeling and Simulation: Applications, Methodology, Technology. 2012;9(3):195–203.

  At its core, cyber SA requires perceiving, encoding (representing), and understanding the environment in terms of how information, events, and actions will impact goals and objectives, both now and in the near future. This study models the cognitive layer of SA by defining a statistical behavioral model that is contingent on user activities. This model provides a representation of normal user behavior based on activities. Hence, deviations from the model prediction can be flagged as suspicious network activities.

- Chen P-C et al. Experience-based cyber situation recognition using relaxable logic patterns. Proceedings of the 2012 IEEE International Multi-Disciplinary Conference on Cognitive Methods in Situation Awareness and Decision Support; 2012.

  The authors developed a systematic approach to leverage experiences of defenders to enhance cyber situation recognition. Using a logic-based approach, they efficiently capture and use defenders' experiences, which are the categorized as kind of knowledge-based intrusion detection.

- Yen J et al. RPD-based hypothesis reasoning for cyber SA. In: Cyber situational awareness. New York (NY): Springer US; 2010. p. 39–49.

  The paper describes a high-level, cognitively inspired framework that is built upon a Recognition-Primed Decision model, and integrates the 3 components of the traditional SA model. The framework connects the logic world of tools for cyber SA with the mental world of human analysts, enabling the perception, comprehension, and prediction of cyber situations for better prevention, survival, and response to cyber attacks by adapting missions at the operational, tactical, and strategic levels. Further development is needed to formalize an executable computational model build based on this framework.

- Domínguez IX et al. Human subtlety proofs: using computer games to model cognitive processes for cybersecurity. International Journal of Human-Computer Interaction; 2016 Oct 3. doi.org/10.1080/10447318.2016.1232229.

  The paper provides a brief overview of work in different areas of cyber-SA where cognitive modeling research plays a role with regard to direct interaction between end users and computer systems and with regard to the needs of security analysts working behind the scenes. The authors address the fundamental challenge of confirming whether the entity who triggered an activity in the cyber space is human or not (i.e., a software bot). The authors demonstrate that by calibrating cognitive models to human behavior it is possible to characterize human behavior and use it for generating human interactive proof for bot detection.

- Greitzer FL, Hohimer RE. Modeling human behavior to anticipate insider attacks. Journal of Strategic Security. 2011;4(2):25.

  The authors propose a comprehensive threat assessment approach that will provide automated support for the detection of high-risk behavioral "triggers" to help focus the analyst's attention and inform the analysis when detecting insider threat. This predictive modeling framework integrates a diverse set of data sources from the cyber domain as well as inferred psychological/motivational factors that may underlie malicious insider exploits. The main components of the framework are 1) ontologies, representing specialized domain knowledge, 2) reifiers, used for the ingesting of the primitive data types that are specified in the domain ontologies, 3) memory storing facts asserted from the primitive data and the facts inferred by the reasoning system, and 4) reasoning components, used to interpret the data assertions and infer new assertions.

- Siraj A, Vaughn RB. A cognitive model for alert correlation in a distributed environment. Proceedings of the International Conference on Intelligence and Security Informatics. Berlin: (Germany): Springer; 2005.

  In this paper the authors demonstrate the use of cognitive modeling of cause and effect relationships to correlate alerts in a distributed environment. By combining a casual knowledge-based inference technique together with fuzzy cognitive modeling, the authors were able to discover casual relationships in alert data in multiple datasets of cyber exercises.

### 4.3 Learning

Training defenders and their learning to detect new threats based on past experiences are 2 areas that can significantly benefit from modeling and simulating human cognition. Understanding the cognitive mechanisms that govern human adaptation to changes in the highly dynamic cyber environment is a key factor to threat and anomaly detection. The following papers study these issues using laboratory experimentation with a simplified Intrusion Detection System:

- Dutt V et al. Modeling the effects of base-rates on cyber threat detection performance. Proceedings of the 11th International Conference on Cognitive Modeling. Berlin (Germany): Universitaetsverlag der TU; 2012.

  This paper used a cognitive model of a cyber defender to study memory retrieval, decision making, and learning from experience. More specifically, the model allowed examination of how defenders perceive and attend situational attributes. The cognitive model accounted for changes in defender's performance when encountering different adversarial behavior.

- Dutt V et al. Cyber situational awareness modeling detection of cyber attacks with instance-based learning theory. Human Factors: The Journal of the Human Factors and Ergonomics Society. 2013:55(3):605–618.

  This paper examined through cognitive modeling how training, risk taking, and past experiences influence the defender's judgment and decision making. Based upon model predictions, a defender's prior threat experiences and tolerance to threats are likely to predict detection accuracy. In parallel, considering the nature of adversarial behavior, especially attacks base rate, is equally important.

## 4.4 Modeling Team Interactions and Decision Making

Cyber defenders often work in heterogeneous and interdependent teams of individuals. Such teams plan, decide, perceive, design, solve problems, and act as an integrated system. Achieving and maintaining effective teamwork heavily depends on communication, information sharing, and collaboration. Typically, team members differ in their areas of expertise and their familiarity with different aspects of the network. Team interactive cognition is a key component to understanding team performance and synergistic behavior. Network security live exercises like Capture the Flag provide rich data for this line of research. The following papers focus on modeling and simulating cyber interactions at the team level:

- Rajivan P et al. Agent-based model of a cyber security defense analyst team. Proceedings of the Human Factors and Ergonomics Society Annual Meeting. 2013;57(1).

  This study explored the effects of different collaboration strategies and team sizes on performance measures such as number of intrusion alerts accurately processed by the defenders and rewards they accrue from accurately processing the alerts. This study also explored the feasibility of using agent-based modeling methodologies for studying team processes in the cyber defense context. The model revealed that specific collaboration strategies lead to better performance and that large teams are detrimental to performance.

- Kotenko I. Agent-based modeling and simulation of cyber warfare between malefactors and security agents in Internet. Presented at the 19th European Simulation Multiconference; 2005.

  The author used teams of defender agents and attacker agents to simulate the dynamics in a Distributed Denial of Service attack. The paper presents the structure of the teams of agents, specifications of hierarchies of agent plans, agent interaction and coordination mechanisms, and agent role-assignment mechanisms.

- Ben-Asher N, Gonzalez C. CyberWar game: a paradigm for understanding new challenges of cyber war. In: Cyber warfare. Berlin (Germany): Springer; 2015. p. 207–220.

  This study examines the decision-making processes that drive the dynamics of cyber war using a multi-agent model comprising cognitive agents that learn to make decisions from experience. In this paradigm—the CyberWar

game—assets and power are 2 key attributes that influence the decisions of agents. Assets represent the key resource that an agent is protecting from attacks, while power represents technical prowess of an agent's cyber security. All of the agents share the same goal of maximizing their assets, and they learn to attack other agents and defend themselves to meet this goal. These agents do not learn by using predefined strategies, as many multi-agent models do, but instead learn from experience according to the situation and actions of others, as suggested by many cognitive architectures.

## 5. Conclusion and Recommendations

Multiple disciplines within cyber systems engineering and operations cannot take a mature, rigorous, and suitably advanced form—comparable to those in more mature technical findings—without availability of M&S tools that are specifically targeted for phenomena of cyber defense and produce results at several levels of detail.

- Recommendation: Stress the need for M&S for a full range of cyber specialties—not only for training and rehearsal. We observe very limited availability of commercial products and probably a corresponding lack of commercial research and development.

- Recommendation: Encourage participation of commercial companies, especially developers of products for military command, control, communications, computers, intelligence, surveillance, and reconnaissance requirements, in NATO STO activities and meetings, with an option to demonstrate their relevant products. Several types of simulations are especially relevant to NATO military operations, especially the ability to perform realistic wargaming of a cyber battle—attack and defense—and dynamic replanning for enhanced resilience of NATO systems and networks.

- Recommendation: The types of simulation models that we call here Business Impact Simulation are particularly important for NATO mission planning, mission risk assessment, mission risk mitigation plans, enemy COA assessment, and mission damage assessment. On the other hand, types of simulation models that we call here Attack Details Simulation are crucial for NATO system engineering, model validation, forensic analysis, and generation of indicators.

- Recommendation: Produce a set of clear and concrete requirements for simulation tools specifically targeted at cyber defense mission planning and wargaming for a military unit in active operations against a cyber-capable adversary. This will encourage and guide the development of relevant tools in NATO nations' government and industry.

- Recommendation: Produce a set of clear and concrete requirements for simulation tools capable of simulating attack-defense scenarios at an observable level of network-and-system-level events. Approaches based on attack graphs and related methods appear to be most mature and suitable for certain types of simulation tools. They are likely to be actively pursued by commercial developers.

- Recommendation: Minimize government investments in the line of approaches based on attack graphs and related methods to invest in other directions of cyber M&S. Human (both defenders and attackers) cognitive processes and resulting behaviors are crucial components of simulating a cyber battle. Approaches to such simulations are at the infancy stage.

- Recommendation: NATO nations should encourage academic research targeted at effective and validated simulation of human cognitive processes and behaviors as they execute cyber defense and attack. Two recent large efforts—AMICA and the European Union's Panoptesec—are highly valuable for understanding the state of the art, the gaps, and technical challenges. Lessons of such efforts should be actively studied and used as technical references in developing NATO cyber capabilities. Similarly, the utility of model-based approaches in SE (Estefan 2008) for cyber needs to be considered in view of their greater use in cyber space sectors.

INTENTIONALLY LEFT BLANK.

# Appendix A: Proposal for Workshop IST-156

# TECHNICAL ACTIVITY PROPOSAL (TAP)

| ACTIVITY REFERENCE NUMBER | IST-156 | ACTIVITY TITLE  Modelling and Simulation S&T: Critical Enabler for Cyber Defense | APPROVAL 2016 |
|---|---|---|---|
| TYPE AND SERIAL NUMBER | RWS-022 | | START January 2016 |
| LOCATION(S) AND DATES | | | END January 2017 |
| COORDINATION WITH OTHER BODIES | | HFM, MSG, NCIA, NIAG, ACT | |
| NATO CLASSIFICATION OF ACTIVITY | | Public Release | Non-NATO Invited Yes |
| PUBLICATION DATA | | TR | PR |
| KEYWORDS | Cyber Defense, Modelling and Simulation, Risk, likelihood and Impact Models, Adversary Models, Human Cognition, Resilience Engineering, Graph Theory, Communications and Information Systems, Model Based Systems Engineering, Synthetic environments, Cyber Education and Training | | |

## A-1  Background and Justification (Relevance to the North Atlantic Treaty Organization [NATO])

This technical activity proposal for a workshop addresses the recommendation from the research of the NATO Information Systems and Technology (IST) panel exploratory team, IST-ET-094: "Model-Driven Paradigms for Integrated Approaches for Cyber Defense". Key findings of this research are that the science and technology (S&T) underpinning the developments of models, simulators/emulators, methods, and tools in support of an integrated cyber defense approach remain immature. Though some progress has been made in a few individual cyber-related topic areas (e.g., emulation for cyber ranges and training environments as well as attack graphs analysis), there is less evidence of integration and an interdisciplinary approach to address challenges in cyber defense, a priority strategic risk area for NATO and most, if not all, of its coalition nations.

It is therefore argued that a new model-driven paradigm, interdisciplinary and coherent by design to link the technical areas, is needed for cyber defense to assist the integration of business or military operations with cyber defense, particularly in bridging the cognitive gap between operational decision makers and cyber defenders. This paradigm should support the ability, at the required level of abstraction, to translate operational priorities into cyber defense priorities and to translate impacts on the supporting cyber infrastructure into consequences expressed in operational terms.

## A-2  Objectives

The objectives of the workshop are to provide such a forum and consider the S&T developments (what?, where?, and how?) that could be leveraged and form an integral part of a model-driven approach to arrive at a better representation, with varying levels of fidelity, of the socio-technical system (of systems) that comprise cyber security and defense against cyber threats. This includes the underpinning analytics to determine, with certain levels of validation or assurance, the effects/impacts of threats, options, and consequences of mitigations and actions. This needs outputs and evidence from a broad range of analysis activities: detecting attacks in a mission-supporting manner, assessing damages relevant to the mission, investigating impacts on mission elements, recovering from attacks in order to continue missions to the maximum extent possible, and deciding on how to respond to cyber attacks in a manner that maximizes mission success. Additionally, forensics methods and tools are necessary to determine key facts relevant to assessing mission impact. Such tools are used for evidence collection, analysis of the attack, identification of the attacker, understanding the attack, damage assessment, and attribution of attackers. Depending on the mission and the type of an attack, there may be different degrees of relative importance and resources attached to attack detection, continuity of the mission, damage assessment, evidence collection, attribution, and other activities. Use of related methods, procedures, tools, or technology should depend largely on the mission.

The following 4 themes emerge to provide focus for the workshop for modeling and simulation within this overall cyber defense context:

- Model-Based Systems Engineering (MBSE) approaches: From MBSE developments in other sectors (e.g., land, air, and space), which methods, tools, and techniques can be best applied to socio-technical systems (of systems) that define cyber, their utility, and measures of effectiveness? In particular, how well can a MBSE approach address all layers from the cognitive and virtual to the physical as well as over the system life cycles?

- Models for and simulation of attack (and defend) processes: What can be identified and inferred at a detailed level of observable events (e.g., system logs at lower levels) suitable for the validation of the model and verification of defensive measures? Important aspects include comparison with real events (threats and countermeasures where available), forensics, and warnings and indicators.

- Better representation of modeling and sim(em)ulation of adversarial interactions between attackers and defenders and adversaries themselves: What are the effective high-level techniques and detailed game theoretic techniques for "war gaming" of a range of simple-to-complex scenarios in order to provide means by which solutions can be explored in an overall end-to-end chain of actions-counteraction in an attack or goal-compromise episode?

- Modeling and sim(em)ulation of human cognition and behaviors: How can we model and "encode"' human decision making and action/reaction of a human defender, including working with machines with increasing levels of autonomy, as a function of the environments, the situation at hand, different stimuli, knowledge, experience, and training?

The expected outcome of the workshop is deeper insight into the S&T art of the possible, now and in the future, of a model-based systems approach applied to cyber defense—the utility, benefits, and challenges. Specific technical challenges that need to be addressed to extract value for enhanced cyber defense capability of NATO and coalition nations. In addition, also expect some foresight on how the 2 areas may evolve in the future and the opportunities and threats posed.

### A-3  Topics to Be Covered

- Mature MBSE approaches for cyber defense

- Enhanced models for and simulation of attack (and defend) processes

- Better representation of modeling and sim(em)ulation of adversarial interactions between attackers and defenders, plus adversaries themselves

- Effectiveness in the modeling and sim(em)ulation of human cognition and behaviors

### A-4  Deliverable (e.g., S/W Engage Model, Database,…) and/or End Product (e.g., Final Report)

Technical Report; other deliverable(s): none

Technical Team Leader: Co-Chair: Mr Jack Branhill, United Kingdom

Lead Nation: United States

### A-5  Nations Willing/Invited to Participate

NATO nations and bodies: Belgium, Bulgaria, France, Germany, Portugal, Romania, the United Kingdom, and the United States

Partnership for Peace nations: Finland, Montenegro, and Sweden

Mediterranean Dialogue nations: none

Istanbul Cooperation Initiative nations: none

Global Partners: Australia

Contact/other nations: none

### A-6  National and/or NATO Resources Needed (Physical and Nonphysical Assets)

Nations are asked to provide, from their cyber community, leading subject matter experts and suitable representatives from their stakeholders, policy makers, and practitioners, including new STEM (Science, Technology, Engineering and Mathematics) staff. A background and experience, current or latent, in the 4 topics detailed in Section C-3 and operational experience in implementing cyber defense solutions is very desirable.

## A-7  STO/CSO Resources Needed

Provision of funding for and support for the arrangements of a technical evaluator for the workshop and 2 keynote speakers.

# Appendix B: Special Issue of the *Journal of Defense Modelling and Simulation* (JDMS)

# Call for Papers

*Journal of Defense Modeling and Simulation*:
Applications, Methodology, Technology (JDMS)

**Special Issue:** Model-Driven Paradigms for Integrated Approaches to Cyber Defense

**Guest Editors**

Dr Alexander Kott, US Army Research Laboratory, Adelphi, Maryland

**Introduction**

The growing military importance of cyber security is unquestionable. Increased use of commercial off-the-shelf information technology and dependency on computerized information systems for weapons, intelligence, communication, and logistics continues to increase vulnerability of military missions to cyber attacks. Successful mission execution requires highly capable technologies that result in forces performing a broad range of defensive cyber operations for each step of an attacker's life cycle.

While many of these defensive cyber operations rely on essentially ad hoc techniques, there is a growing realization within the cyber community that a comprehensive, systematic, principle-based modeling and simulation approach is more likely to produce long-term, lasting, and reusable approaches for defensive cyber operations. Such a paradigm is predicated on the creation and validation of mechanisms of modeling the organization whose mission is subject to assessment, the mission (or missions) itself, and the cyber-vulnerable systems that support the mission. The models are then used to simulate or otherwise portray the cyber attacks and associate defensive phenomena and system operations, including the assessment of mission impact.

The main objective of this special issue is to offer the readers a broad, yet integrated exploration of the field while providing a publishing venue for researchers working toward a multipurpose, integrated system of cyber models that guide a broad range of cyber security operations, examples of which include vulnerability analysis, intrusion prevention, intrusion detection, analysis, forensics, attribution, mission impact assessment, and recovery.

Candidate model-driven paradigms questions for cyber defense include the following. Are there applications of the model-driven paradigm that are more likely to prove fruitful in the near term than others? What can be learned and adopted from the ongoing efforts, such as experiences in the European Union Panoptesec program that explores a model-based approach? What are ways to populate and validate models in an affordable fashion? Is the model-driven paradigm defeated by ever-growing diversity and diffusion of information technology infrastructures, such as Internet of Things? What commercial tools are emerging that can support the model-driven paradigm? Could these approaches be adapted for military-specific requirements?

Possible topics for authors to consider include the following:

- Theoretical foundations and formulations of the model-driven paradigm
- Relevance of game-theoretic and control-theoretic approaches
- Formal languages for model specification
- Assessment of barriers to successful use of the model-driven paradigm
- Potential techniques for using model-driven paradigms for cyber defense problem solving at different phases of cyber operations (e.g., prior, during, and after discovery of a cyber compromise)
- Analysis of known related approaches and methods
- Complexity and completeness of the models
- Feasibility of automated or semi-automated generation of models
- Modeling of the adversary
- Human factors in the models
- Calibration of the models
- Validation of the models
- Maintenance of the models
- Utility functions to be used in conjunction with models

Papers submitted should not be concurrently under review at another conference, journal, or similar venue.

**Instructions for Manuscript Preparation**

For manuscript formatting and other guidelines, visit the Author Guidelines for JDMS. Manuscripts must not have been previously published or be submitted for publication elsewhere. Each submitted manuscript must include title, names, authors' affiliations, postal and e-mail addresses, and a list of keywords. For multiple author submission, please identify the corresponding author.

**Due Dates**

Submission of papers: September 30, 2016

Expected date of publication: Summer 2017

**Guest Editors**:

- Alexander Kott, PhD

    Chief of Network Science Division, US Army Research Laboratory, Adelphi, Maryland, USA

- Nazife Baykal, PhD

    Professor, Director of Informatics Institute and Chair, Cyber Security Department, Middle East Technical University, Ankara, Turkey

- Yilmaz Cankaya

    Chief Researcher, TÜBITAK BILGEM Cyber Security Institute, Kocaeli, Turkey

- Bob Madahar, PhD

    Professor, Senior Fellow, DSTL, Porton, United Kingdom

- Col. Nikolai Stoianov, PhD

    Associate Professor, Defence Institute "Prof. Tsvetan Lazarova", Sofia, Bulgaria

- Margaret Varga, PhD

    Director at Seetru Ltd, visiting fellow at the University of Oxford. Bristol, United Kingdom

**For questions contact:**

Vicki Pate, Managing Editor \Journal of Defense Modeling & Simulation: vmpate@scs.org

Dr Alexander Kott, Guest Editor: alexander.kott1.civ@mail.mil

# Appendix C: Proposal for Exploratory Team ET-094

# TECHNICAL ACTIVITY PROPOSAL (TAP)

| ACTIVITY REFERENCE NUMBER | IST-___ | ACTIVITY TITLE | APPROVAL 2015 |
|---|---|---|---|
| TYPE AND SERIAL NUMBER | ET-___ | Model-Driven Paradigms for Integrated Approaches to Cyber Defence | START December 2015 |
| LOCATION(S) AND DATES | | TBD | END July 2016 |
| COORDINATION WITH OTHER BODIES | | NATO ACT, NCIA, CCDCOE | |
| NATO CLASSIFICATION OF ACTIVITY | | NATO UNCLASSIFIED, RELEASABLE TO PUBLIC | Non-NATO Invited Yes |
| PUBLICATION DATA | | R | NU |
| KEYWORDS | | Cyber Defense Operations, Computer Information Systems (CIS), Information & Communications Technology (ICT), Modeling and Simulation | |

## C-1 Background and Justification (Relevance to North Atlantic Treaty Organization [NATO]):

The growing military importance of cyber space and cyber security is unquestionable. The increased use of commercial off-the-shelf information technology and dependency on computer information systems for weapons, intelligence, communication, and logistics continues to increase vulnerability of military missions to cyber attacks. Successful execution of NATO missions requires highly capable technologies that allow the forces to perform a broad range of defensive cyber operations, including vulnerability identification, risk characterization, continuous monitoring, evidence collection, analysis of the attack, identification of the attacker, understanding the attack, forensic analysis, damage assessment, mission impact assessment, attribution of attackers, threat analysis, remedial stabilization of the system, active response to attach and preventive actions.

While much of these defensive activities currently rely on essentially ad hoc techniques, there is a growing realization within the cyber community that comprehensive, systematic, principle-based modeling and simulation are more likely to produce long-term, lasting, reusable approaches to defensive cyber operations. In particular, a recent NATO workshop (IST-128) voiced a strong consensus that substantive solutions for at least some classes of cyber challenges, for example mission impact assessment, was in adopting and developing a new model-driven paradigm.

Such a paradigm is predicated on creation and validation of mechanisms of modeling the organization whose mission is subject to assessment, the mission (or missions) itself, and the cyber-vulnerable systems that support the mission. The models are then used to simulate or otherwise portray the cyber attacks and associate defensive phenomena and system operations, including the assessment of mission impact. In addition, such model-based analysis could be used to explore multiple alternative mitigation and work-around strategies and select the optimal course of mitigating actions.

Furthermore, the model-driven paradigm applied to cyber operations is likely to benefit traditional disciplines of cyber defense, such as vulnerability analysis, intrusion prevention, intrusion detection, analysis, forensics, attribution, and recovery. For example, intrusion detection, especially for zero-day or polymorphic attacks, would greatly benefit from the ability to model the observable effects of a hypothetic attack.

## C-2 Objective(s):

The main objective of the activity is to perform a broad and unconstrained exploration of the possibility of using a multipurpose, integrated system of models for guiding a broad range of cyber security operations: vulnerability analysis, intrusion prevention, intrusion detection, analysis, forensics, attribution, mission impact assessment, and recovery. The team will investigate whether such a paradigm can be expected to provide meaningful, actionable information about cyber activities and the resulting impacts that have not been seen before or do not match prior experiences and patterns.

To be sure, a model-driven paradigm presents a number of serious challenges both practical and theoretical. For example, a major challenge is to create and maintain an appropriate model. Manual generation of such a complex model appears possible in some cases but prohibitively expensive and impossible to maintain in the long term. Therefore, it will be ET-094's charter to explore such challenges and answer a number of questions, such as the following: How likely these challenges be overcome in the near- and mid-term? Are there applications of the model-driven paradigm that are more likely to prove fruitful in near-term than others? What can be learned and adopted from the ongoing efforts, such as experiences in the European Union Panoptesec program that explore a model-based approach? What are ways to populate and validate models in an affordable fashion? Is the model-driven paradigm defeated by ever-growing diversity and diffusion of information technology infrastructures, such as the Internet of Things? What commercial tools are emerging that can support the model-driven paradigm? Could these be adapted for military-specific requirements?

## C-3 Topics to Be Covered

- Theoretical foundations and formulations of the model-driven paradigm
- Relevance of game-theoretic and control-theoretic approaches
- Formal languages for model specification
- Assessment of barriers to successful use of the model-driven paradigm
- Potential techniques for using model-driven paradigm for cyber defense problem solving at different phases of cyber operations (e.g., prior, during, and after discovery of a cyber compromise)
- Analysis of known related approaches and methods
- Complexity and completeness of the models

- Feasibility of automated or semi-automated generation of models
- Modeling of the adversary
- Human factors in the models
- Calibration of the models
- Validation of the models
- Maintenance of the models
- Utility functions to be used in conjunction with models

## C-4  Deliverables and/or End Product

- Final Report
- Collection of materials (briefs, papers, etc.) assembled in support of the Final Report
- Recommendations regarding the model-driven paradigms for cyber defense, including topics for future activities if appropriate

## C-5  Technical Team Leader and Lead Nation:

Chair:          TBD

Lead Nation:    TBD

## C-6  Nations Willing/Invited to Participate:

Poland, Rumania, Turkey, Germany, United States

Partnership for Peace (PfP) nations: all PfP invited

Mediterranean Dialogue nations: none

Istanbul Cooperation Initiative nations: none

Global Partners: none

Contact/Other nations: none

## C-7  National and/or NATO Resources Needed (Physical and Nonphysical Assets)

- Support from national host for meeting

## C-8  RTA Resources Needed (e.g., Consultant Funding)

- None

# Appendix D: Skybox Security Software

After reviewing commercial off-the-shelf solutions, we identified Skybox Security as a commercially available cyber security management software. Skybox was contacted and agreed to organize a presentation in April 2016. The following is a summary of findings resulting from this presentation.

Their solution seems quite advanced in terms of network discovery, reachability map reconstruction, and capability to manipulate this reconstruction via a graphical user interface (GUI). The tool is able to use different vulnerability scanners to extract vulnerability data. Skybox uses a Panoptesec-similar algorithm for the computation of the attack paths, based on the Common Vulnerabilities and Exposures system (https://cve.mitre.org/), and proactively analyses the network to propose mitigation actions for risk reduction. While Skybox's cyber security management software is similar to the Panoptesec approach, it does not take into account business-related analysis. It is possible to define some particular devices as critical, to avoid or optimize the instantiation of the mitigation actions on them, but no more than that. Skybox completely avoids the reactive perspective, and the company is actually seeking and developing its reactive approach. Skybox seems to be quite strong on the configuration/GUI and seems able to manage thousands of devices, which means their attack path algorithms are quite optimized. The interesting discovery is the lack of an evolved business mission engine. The engineer who made the demo for me seemed to understand my questions about such an engine, but I am not sure it is in the development plan of the company to go much deeper in this direction (compared with the kind of evolution we envisioned for our exploratory team). Skybox, in conclusion, is not particularly advanced in terms of the concepts we analyzed and apparently not extremely interested in them.

## List of Symbols, Abbreviations, and Acronyms

| | |
|---|---|
| AMICA | analyzing mission impacts of cyber actions |
| CAMIAC | Cyber Attack Modeling and Impact Assessment Component |
| COA | course of action |
| CR | cyber range |
| CySeMoL | Cyber Security Modeling Language |
| GUI | graphical user interface |
| IST | Information Systems and Technology |
| IT | information technology |
| JDMS | *Journal of Defense Modelling and Simulation* |
| M&S | modeling and simulation |
| MAST | Malicious Activity Simulation Tool |
| MBSE | Model-Based Systems Engineering |
| NATO | North Atlantic Treaty Organization |
| PfP | Partnership for Peace |
| S&T | science and technology |
| SA | situational awareness |
| SE | Systems Engineering |
| SIR | Susceptible, Infected, Recovered |
| STO | Science and Technology Organization |
| UK | United Kingdom |

1    DEFENSE TECH INFO CTR
 (PDF)  DTIC OCA

    2    US ARMY RSRCH LAB
 (PDF)  IMAL HRA MAIL & RECORDS MGMT
           RDRL CIO L TECHL LIB

    1    GOVT PRNTG OFC
 (PDF)  A  MALHOTRA

    1    US ARMY RSRCH LAB
 (PDF)  RDRL CIN
            A  KOTT